\documentclass[conference]{IEEEtran}
\usepackage{fancyhdr}
\IEEEoverridecommandlockouts
\usepackage[hidelinks]{hyperref}
\usepackage{cite}
\usepackage[T1]{fontenc}
\usepackage{amsmath,amssymb,amsfonts}
\usepackage{graphicx}
\usepackage{textcomp}
\usepackage[table]{xcolor}
\usepackage{float}
\usepackage{orcidlink}
\usepackage{xurl}
\usepackage{siunitx}
\DeclareSIUnit{\flop}{FLOP}

\usepackage{algorithm}
\usepackage{algpseudocode}
\usepackage{tikz}
\usetikzlibrary{arrows.meta, patterns}
\usepackage{caption}
\usepackage{tabularx}
\usepackage{booktabs}
\usepackage[acronym,nomain]{glossaries}

\makeglossaries
\newacronym[longplural={Gaussian processes}]{gp}{GP}{Gaussian process}
\newacronym{gpr}{GPR}{Gaussian process regression}
\newacronym{hpc}{HPC}{high-performance computing}
\newacronym{blas}{BLAS}{Basic Linear Algebra Subprograms}
\newacronym{lapack}{LAPACK}{Linear Algebra PACKage}
\newacronym{bbmm}{BBMM}{Blackbox Matrix-Matrix multiplication}
\newacronym{msd}{MSD}{mass-spring-damper}
\newacronym{usm}{USM}{unified shared memory}
\newacronym{ml}{ML}{machine learning}
\newacronym{uxl}{UXL}{Unified Acceleration}
\newacronym{ai}{AI}{Artificial Intelligence}

\colorlet{potrf}{purple}
\colorlet{trsm}{teal}
\definecolor{syrk}{rgb}{0.8, 0.8, 0}
\colorlet{gemm}{blue!65}
\definecolor{utile}{rgb}{0.8, 0.8, 0.8}

\newcolumntype{P}[1]{>{\centering\arraybackslash}p{#1}}
\newcolumntype{M}[1]{>{\centering\arraybackslash}m{#1}}

\begin{document}
\bstctlcite{IEEEexample:BSTcontrol}

\title{Performance-Portable Exact Gaussian Process Regression in Python via SYCL and oneMath}
\author{
	\IEEEauthorblockN{1\textsuperscript{st} Alexander Strack\orcidlink{0000-0002-9939-9044},
		2\textsuperscript{nd} Marcel Graf\orcidlink{0009-0007-6308-1329},
		3\textsuperscript{rd} Alexander Van Craen\orcidlink{0000-0002-3336-7226},
		4\textsuperscript{th} Dirk Pfl\"uger\orcidlink{0000-0002-4360-0212}}
	\IEEEauthorblockA{\textit{IPVS, University of Stuttgart}\\
		Stuttgart, Germany \\
		alexander.strack@ipvs.uni-stuttgart.de, st172528@stud.uni-stuttgart.de, \\
		alexander.van-craen@ipvs.uni-stuttgart.de, dirk.pflueger@ipvs.uni-stuttgart.de}
}

\maketitle
\thispagestyle{fancy}
\lhead{}
\rhead{}
\chead{}
\lfoot{\parbox{\textwidth}{\scriptsize\raggedright
This is the accepted version of a paper to appear in the Proceedings of the SC26 Workshops (WACCPD 2026), November 15--20, 2026, Chicago, Illinois, USA. The final version will be available via IEEE Xplore.\\[2pt]
\copyright{} 2026 IEEE. Personal use of this material is permitted. Permission from IEEE must be obtained for all other uses, in any current or future media, including reprinting/republishing this material for advertising or promotional purposes, creating new collective works, for resale or redistribution to servers or lists, or reuse of any copyrighted component of this work in other works.}}
\rfoot{}
\cfoot{}
\renewcommand{\headrulewidth}{0pt}
\renewcommand{\footrulewidth}{0pt}

\begin{abstract}
	Modern multi-vendor GPU environments force a choice between portable code and native performance.
	GPRat is an open-source Python library for exact Gaussian process regression built on the HPX asynchronous many-task runtime, tiled algorithms, and CPU/CUDA BLAS backends.
	We contribute a UXL Foundation oneMath backend with a hybrid HPX/SYCL design, extending GPRat to AMD and Intel GPUs.
	The new SYCL backend comes within \qty{3}{\percent} of native CUDA on an NVIDIA A30 at the largest problem size and outperforms the GPflow and GPyTorch libraries by up to $1.75\times$ and $2.17\times$ on the A30 and up to $1.79\times$ and $4.62\times$ on an AMD Instinct MI210.
	On an Intel Arc B580, a consumer-class card with FP64 throughput below our CPU's, the unmodified SYCL backend trails GPRat's CPU backend.
	Overall, oneMath's dispatch to vendor-native BLAS lets a single SYCL codebase match CUDA on NVIDIA and outperform established Python GP frameworks on AMD, without vendor-specific code.
\end{abstract}


\section{Introduction}
More than half of the systems on the TOP500 list of the world's fastest supercomputers use accelerators, predominantly manufactured by NVIDIA, AMD, or Intel~\cite{top500}.
Yet the dominant GPU programming models remain vendor-specific.
Writing and maintaining a separate native-code path per vendor is costly, so a single portable implementation is attractive, provided its performance overhead relative to native code is small and well quantified.
SYCL~\cite{SyclSpec}, backed by vendor-native math libraries through oneMath~\cite{OneMathSpec}, is a candidate for closing this portability gap, and measuring the resulting portability overhead is a central concern of this work.

We study this portability question in the setting of \glspl{gp}.
\Glspl{gp} are a non-parametric alternative to the parametric models, such as neural networks, for various \gls{ml} applications~\cite{Rasmussen2005_gp_ml}.
Unlike neural network models, a \gls{gp} naturally provides uncertainty estimates for its predictions and allows the direct integration of prior knowledge, such as periodicity~\cite{Kocijan2016_gp_si_book}.
\glspl{gp} are applied, inter alia, in control theory~\cite{Kocijan2016_gp_si_book}, geostatistics~\cite{Mackay1998_intro_gp, Williams1995_gpr}, \gls{ai}~\cite{Damianou2013_deep_gps}, scientific computing~\cite{Kennedy2001_gp_sc}, big data~\cite{Liu2020_gp_bigdata}, chemistry~\cite{Deringer2021_gpr_chemistry}, and psychology~\cite{Schulz2018_gpr_tutorial}.

Established \gls{gpr} Python libraries largely inherit their portability from a general-purpose backend.
As native Python implementations struggle to harness modern \gls{hpc} devices~\cite{Castro2023_python_slow}, many libraries write their core logic in C or C\texttt{++} and expose it through Python bindings.
Popular examples in the context of \gls{ml} include PyTorch~\cite{Paszke2019_pytorch} and TensorFlow~\cite{Abadi2016_tensorflow}.
GPyTorch~\cite{Gardner2018_gpytorch} and GPflow~\cite{Matthews2017_gpflow} are two notable \gls{gpr} implementations built on top of these general-purpose libraries.
Both focus on a broad feature set and delegate performance, scalability, and hardware portability to their respective backends, which can be a limitation when targeting heterogeneous hardware.
Moreover, on less common hardware, installing GPyTorch and, especially, GPflow can be complicated due to strict and sprawling dependencies.

To address these challenges, we developed the GPRat~\cite{Helmann2026_gprat} library.
It provides a Python interface via pybind11~\cite{Wenzel2017_pybind11} and, under the hood, leverages asynchronous, task-based parallelization via the HPX~\cite{Kaiser2020_hpx} runtime.
Its entire dependency tree is available via Spack~\cite{Gamblin2015_spack}, which contributes to a mature installation pipeline.
Portability is a core design goal of GPRat, demonstrated across x86, ARM, and RISC-V CPUs~\cite{Strack2026_gprat_riscv} and across NVIDIA accelerators~\cite{Moellmann2026_gprat_gpu}.
Performance and portability across accelerators from both AMD and Intel remained unclear.

In this work, we present the following contributions:
\begin{itemize}
	\item A portable SYCL backend for GPRat providing GPU support across multiple vendors, leveraging oneMath~\cite{OneMath} for vendor-native \gls{blas} performance,
	\item A hybrid HPX/SYCL scheduling design in which HPX orchestrates GPRat's custom per-tile kernels while oneMath \gls{blas} calls run outside the task graph,
	\item An evaluation of the new backend on an NVIDIA A30 GPU, an AMD Instinct MI210 GPU, and an Intel Arc B580 GPU, including a quantification of the portability overhead of SYCL relative to a native CUDA implementation,
	\item A performance comparison against a oneMKL CPU reference and two GPU reference implementations using GPyTorch and GPflow.
\end{itemize}

This work is structured as follows.
First, Section~\ref{sec:related} highlights selected related work.
Then, Section~\ref{sec:stack} introduces the central components of the employed software stack.
Section~\ref{sec:background} provides the necessary background on \glspl{gp} and tiled Cholesky decomposition.
Section~\ref{sec:implementation} details the integration of the SYCL backend into GPRat.
After that, Section~\ref{sec:results} presents and discusses performance results.
Section~\ref{sec:conclusion} concludes the paper and outlines directions for future work.
Lastly, Appendix~\ref{sec:reproducibility} provides information relevant to reproducing the results.

\section{Related Work}
\label{sec:related}
\gls{gpr} is a longstanding and active field of research, with realizations ranging from mathematically precise approaches to fast approximations.
Tiled Cholesky decomposition is an algorithm for which asynchronous parallelization has shown significant performance improvements~\cite{Strack2026_fork_join}.
Large-scale \gls{gpr} via tiled Cholesky factorization is represented most directly by ExaGeoStat~\cite{Abdulah2018_exageostat_starpu_cholesky}, which targets geostatistical applications using tile low-rank Cholesky on CPU/GPU clusters via the StarPU~\cite{Augonnet2011_star_pu} runtime.
A subsequent PaRSEC-based~\cite{Bosilca2013_parsec} reimplementation added mixed-precision tile Cholesky computations for the same geostatistical workload~\cite{Abdulah2022_parsec_geostat_mixed_precision}.
StarPU and HPX both express computation as task graphs but differ in their scheduling philosophies: StarPU centers on a codelet-and-dependency model with performance-model-driven scheduling, whereas HPX builds a futurized dataflow from standard-conforming C\texttt{++} primitives and uses lightweight user-space threads with work stealing.
GPRat follows the latter approach, and in this work, we further combine HPX dataflow with SYCL runtime scheduling for the GPU \gls{blas} path.

The portability overhead of SYCL relative to native CUDA or HIP has been studied across several application domains, generally reporting modest slowdowns.
In bandwidth-bound cases, SYCL has even been shown to match or slightly exceed native performance when dispatching to vendor-native code paths~\cite{Reguly2023_sycl_portability, Alekseenko2024_gromacs_sycl, Breyer2022_plssvm_sycl_portability}.
Closer to our setting, a SYCL-dispatched oneMKL \gls{blas} backend was shown to match native cuBLAS and rocBLAS performance for moderate and large problem sizes, with overheads observed only for very small operations~\cite{Alpay2022_adaptivecpp}.
For high-level programming models rather than hand-written kernels, Julia and Kokkos implementations were benchmarked on CPU, NVIDIA A100, and AMD MI250X targets against native CUDA and HIP baselines, mostly finding competitive performance, though with gaps on the A100~\cite{Godoy2023_portability_julia_kokkos}.
Outside the \gls{gp} setting, task-based tiled Cholesky decomposition has separately been made portable across NVIDIA and AMD GPUs through the IRIS runtime's multi-vendor dispatch layer~\cite{Monil2024_matris_cholesky}.
OmpSs-2 folds device operations directly into the task graph, letting an OpenACC-managed queue satisfy task dependencies through asynchronous finalization~\cite{Korakitis2022_ompss2_openacc}. The Octo-Tiger CUDA/SYCL executors keep this coupling within HPX by returning a future per device operation~\cite{Daiss2023_hpx_kokkos_sycl}. We take the opposite approach: HPX owns the host-side task graph and custom SYCL kernels while oneMath calls are scheduled by a separate queue pool (Section~\ref{subsec:data-dependency-management}).
Our work contributes a further data point in the \gls{gpr} setting, where the heavy lifting is performed by oneMath-dispatched \gls{blas} kernels rather than hand-written SYCL.

\section{Software Stack}
\label{sec:stack}
The central software components used in this work are the HPX runtime system~\cite{Kaiser2020_hpx}, the Intel oneAPI DPC\texttt{++} implementation~\cite{IntelDpcppCompiler} of the SYCL standard~\cite{SyclSpec}, and the \gls{uxl} Foundation oneMath implementation~\cite{OneMath} of the oneAPI Math Library~\cite{OneMathSpec} standard.

\subsection{HPX}

HPX~\cite{Kaiser2020_hpx} is an asynchronous many-task runtime with a standard-conforming C\texttt{++} interface.
GPRat expresses its tiled algorithms as futurized dataflow.
A tile operation starts as soon as its inputs are ready rather than waiting on a fork-join synchronization~\cite{Strack2026_fork_join}.

Tasks are launched with \texttt{hpx\allowbreak::\allowbreak async} and \texttt{hpx\allowbreak::\allowbreak dataflow}, and dependencies are carried by \texttt{hpx\allowbreak::\allowbreak future} and \texttt{hpx\allowbreak::\allowbreak shared\_future} objects: a task suspends until the futures it depends on become available.
HPX's work-stealing schedulers may then resume a suspended task on a different OS thread than the one it started on.
This thread migration can cause issues when targeting accelerators and is the main reason for our hybrid HPX/SYCL approach.

HPX provides experimental executors for cross-operation with CUDA and SYCL, which were developed for the Octo-Tiger application~\cite{Daiss2023_hpx_kokkos_sycl}.
However, they do not cover certain aspects relevant to GPRat's GPU backends, which is why the CUDA executors were not used in~\cite{Moellmann2026_gprat_gpu}.
Likewise, we do not use the corresponding SYCL executors in this work.

\subsection{SYCL}
\label{subsec:sycl}
SYCL is a programming model defined by the Khronos Group~\cite{SyclSpec}.
It acts as an abstraction layer for heterogeneous computing on various devices.
Multiple implementations in the form of compatible compilers exist, including the \emph{Intel oneAPI DPC\texttt{++}/C\texttt{++} Compiler}~\cite{IntelDpcppCompiler}, also referred to as \texttt{icpx}, which is used in this work.
AdaptiveCpp~\cite{Alpay2022_adaptivecpp}, developed at Heidelberg University, is a notable open-source alternative.
We use \texttt{icpx} because oneMath requires it for Intel GPU support, and using a single compiler across all target vendors simplifies the build infrastructure and improves comparability.
oneMath's current compiler support matrix~\cite{OneMath} illustrates why a single compiler is preferable: AdaptiveCpp is not supported on AMD and NVIDIA GPUs for the \gls{lapack} domain, which provides the \textcolor{potrf}{POTRF} routine central to GPRat's Cholesky decomposition, and Intel GPUs require \texttt{icpx} regardless.

The SYCL standard defines an extensive interface for scheduling kernels and managing data on target devices like CPUs, GPUs, and FPGAs.
Kernels can be written over a \texttt{sycl\allowbreak::\allowbreak range}, a \texttt{sycl\allowbreak::\allowbreak nd\_range}, or hierarchically.
Memory can be managed through a buffer-accessor model or through \gls{usm}.
\Gls{usm} offers \texttt{sycl\allowbreak::\allowbreak malloc\_shared} as well as explicit allocation with \texttt{sycl\allowbreak::\allowbreak malloc\_host} and \texttt{sycl\allowbreak::\allowbreak malloc\_device}.
Our SYCL backend relies on the latter.

Kernels and memory interactions are submitted to SYCL queues.
A queue has a context and targets exactly one device, although multiple queues can target the same device.
The targeted device type can be set by selecting it with the device selector, for instance, targeting a GPU with \texttt{sycl\allowbreak::\allowbreak gpu\_selector\_v}.
By default, all queues operate out of order, so the runtime may launch independent kernels in a different order than they were submitted.

\subsection{UXL Foundation oneMath}

The \gls{uxl} Foundation released a specification for a unified interface to mathematical routines available across a wide range of target devices, known as the \emph{oneAPI Math Library}~\cite{OneMathSpec}.
Together with the aforementioned Intel compilers and further libraries, it belongs to the Intel oneAPI framework~\cite{oneapi}.
For brevity, the standard is also referred to as \emph{oneMath}.

In this work, we use the \gls{uxl} Foundation's open-source implementation of the oneMath specification~\cite{OneMath}.
In the following, any mention of oneMath refers to the implementation used here, not the standard.
It supports NVIDIA, AMD, and Intel GPUs as well as multiple CPU architectures.
In the background, generic calls are routed to the appropriate operation on the selected backend.
For instance, \gls{blas} operations exposed by oneMath are mapped to cuBLAS~\cite{cublas} calls on NVIDIA GPUs.
While release \texttt{v0.9} already exposes both row-major and column-major interfaces, only the latter are implemented.
Row-major storage is the standard convention in C\texttt{++} and Python, and GPRat follows this convention accordingly.
For compatibility with column-major \gls{blas} libraries, mathematical operations are reformulated such that explicit conversion is not necessary~\cite{Moellmann2026_gprat_gpu}.

\begin{figure*}[t]
	\centering
	\begin{minipage}[c]{0.48\textwidth}
		\begin{algorithmic}[1]
			\footnotesize
			\For{$J \gets 0$ \textbf{to} $M - 1$}
			\State \textcolor{potrf}{\textbf{POTRF}}($\mathbf K_{J,J}$)
			\For{$I \gets J + 1$ \textbf{to} $M - 1$}
			\State \textcolor{trsm}{\textbf{TRSM}}($\mathbf K_{J,J}, \mathbf K_{I,J}$)
			\EndFor
			\For{$I \gets J + 1$ \textbf{to} $M - 1$}
			\State \textcolor{syrk}{\textbf{SYRK}}($\mathbf K_{I,J}, \mathbf K_{I,I}$)
			\For{$K \gets J + 1$ \textbf{to} $I - 1$}
			\State \textcolor{gemm}{\textbf{GEMM}}($\mathbf K_{I,J}, \mathbf K_{K,J}, \mathbf K_{I,K}$)
			\EndFor
			\EndFor
			\EndFor
		\end{algorithmic}
		\caption{Tiled right-looking Cholesky decomposition of $\mathbf{K}$, reproduced from~\cite{Moellmann2026_gprat_gpu}.}
		\label{fig:tiled-algo}
	\end{minipage}
	\hfill
	\begin{minipage}[c]{0.48\textwidth}
		\centering
		\begin{tikzpicture}[scale=0.65]
			\fill[utile] (1, 0) rectangle ++(1, -1);
			\fill[utile] (2, 0) rectangle ++(1, -1);
			\fill[utile] (3, 0) rectangle ++(1, -1);
			\fill[utile] (4, 0) rectangle ++(1, -1);

			\fill[utile] (2, -1) rectangle ++(1, -1);
			\fill[utile] (3, -1) rectangle ++(1, -1);
			\fill[utile] (4, -1) rectangle ++(1, -1);

			\fill[utile] (3, -2) rectangle ++(1, -1);
			\fill[utile] (4, -2) rectangle ++(1, -1);

			\fill[utile] (4, -3) rectangle ++(1, -1);

			\fill[potrf] (0, 0) rectangle ++(1, -1);

			\fill[trsm] (0, -1) rectangle ++(1, -1);
			\fill[trsm] (0, -2) rectangle ++(1, -1);
			\fill[trsm] (0, -3) rectangle ++(1, -1);
			\fill[trsm] (0, -4) rectangle ++(1, -1);

			\fill[syrk] (1, -1) rectangle ++(1, -1);
			\fill[syrk] (2, -2) rectangle ++(1, -1);
			\fill[syrk] (3, -3) rectangle ++(1, -1);
			\fill[syrk] (4, -4) rectangle ++(1, -1);

			\fill[gemm] (1, -2) rectangle ++(1, -1);

			\fill[gemm] (1, -3) rectangle ++(1, -1);
			\fill[gemm] (2, -3) rectangle ++(1, -1);

			\fill[gemm] (1, -4) rectangle ++(1, -1);
			\fill[gemm] (2, -4) rectangle ++(1, -1);
			\fill[gemm] (3, -4) rectangle ++(1, -1);

			\draw[step=1cm, white, line width=0.5mm] (0,0) grid (5,-5);

			\node[black] at (0.5, -0.5) {\textbf{P}};

			\node[black] at (0.5, -1.5) {\textbf{T}};
			\node[black] at (0.5, -2.5) {\textbf{T}};
			\node[black] at (0.5, -3.5) {\textbf{T}};
			\node[black] at (0.5, -4.5) {\textbf{T}};

			\node[black] at (1.5, -1.5) {\textbf{S}};
			\node[black] at (2.5, -2.5) {\textbf{S}};
			\node[black] at (3.5, -3.5) {\textbf{S}};
			\node[black] at (4.5, -4.5) {\textbf{S}};

			\node[white] at (1.5, -2.5) {\textbf{G}};

			\node[white] at (1.5, -3.5) {\textbf{G}};
			\node[white] at (2.5, -3.5) {\textbf{G}};

			\node[white] at (1.5, -4.5) {\textbf{G}};
			\node[white] at (2.5, -4.5) {\textbf{G}};
			\node[white] at (3.5, -4.5) {\textbf{G}};

			\foreach \i in {0,...,4} {
					\node[black, font=\footnotesize] at (-0.3, -\i-0.5) {\i};
					\node[black, font=\footnotesize] at (\i+0.5, 0.3) {\i};
				}
		\end{tikzpicture}
		\caption{Example of matrix $\mathbf {K} $ split into $5 \times 5$ tiles, colored by tasks in the first iteration, i.e., $J=0$, reproduced from~\cite{Moellmann2026_gprat_gpu}.}
		\label{fig:tiled-vis}
	\end{minipage}
\end{figure*}

\section{Background}
\label{sec:background}
The following provides essential background knowledge for this work.
First, \glspl{gp} are introduced as a general concept, and \gls{gpr} is presented as one of their applications.
This is followed by an introduction to the tiled algorithm approach to Cholesky decomposition, which is the central feature of GPRat.

\subsection{Gaussian Processes and Regression}
\gls{gpr} refers to using a \gls{gp} as a probabilistic regression model, which is the application this paper and the libraries it evaluates focus on.
The theory behind \gls{gpr} is elaborated thoroughly in~\cite{Kocijan2016_gp_si_book, Rasmussen2005_gp_ml}.
A \gls{gp} prescribes a probability measure over functions $f: \mathbb{R}^D \rightarrow \mathbb{R}$ through the marginalization property: any finite collection of function evaluations follows a multivariate Gaussian distribution.
This is written as
\begin{equation}
	f(\vec{x_i}) \sim \mathcal{GP}(m(\vec{x_i}), k(\vec{x_i}, \vec{x_j}))
\end{equation}
for training samples $x_i,x_j$ where $m(\vec{x_i})$ is a \emph{mean} function, commonly set to zero, and $k(\vec{x_i}, \vec{x_j})$ is a \emph{kernel function} encoding prior assumptions such as smoothness or periodicity~\cite{Rasmussen2005_gp_ml}.

The kernel determines the covariance matrix of observed outputs.
We use the \emph{squared exponential kernel}
\begin{equation}\label{eq:kernel}
	k(\vec{x_i}, \vec{x_j}) = \nu \cdot \exp\left( -\frac{1}{2l^2} \left\Vert\vec{x_i} - \vec{x_j}\right\Vert^2 \right) + \delta_{ij} \sigma^2
\end{equation}
with hyperparameters $\nu$ (vertical scale), $l$ (length scale), and $\sigma^2$ (noise variance), fixed to the values $\nu = 1$, $l = 1$, $\sigma^2 = 0.1$ from~\cite{Moellmann2026_gprat_gpu} for comparability.
In Equation \ref{eq:kernel}, $\delta_{ij}$ refers to the Kronecker delta, which is equal to one when the two indices are the same $(i=j)$ and zero otherwise.
Hyperparameter selection is deferred to future training experiments.

\Glspl{gp} can be used for regression applications where they model an input-output relationship given $N$ training inputs $\mathbf{X} \in \mathbb{R}^{N \times D}$, training outputs $\vec{y}$, and $\hat{N}$ test inputs $\mathbf{\hat{X}} \in \mathbb{R}^{\hat{N} \times D}$.
The dimension $D$ describes the number of training features per sample, also commonly referred to as regressors.
The covariance matrices $\mathbf{K}_{\mathbf{X}, \mathbf{X}}$ (training), $\mathbf{K}_{\mathbf{\hat{X}}, \mathbf{\hat{X}}}$ (test), and cross-covariance $\mathbf{K}_{\mathbf{\hat{X}}, \mathbf{X}}$ are computed from the kernel function.
Assuming zero mean, Bayesian conditioning yields the \emph{predictive mean}
\begin{equation}
	\vec{\hat{y}} = \mathbf{K}_{\mathbf{\hat{X}}, \mathbf{X}}\mathbf{K}_{\mathbf{X}, \mathbf{X}}^{-1}\vec{y}
	\label{eq:predictive-mean-simplified}
\end{equation}
and the \emph{posterior covariance matrix}
\begin{equation}
	\mathbf{\Sigma} = \mathbf{K}_{\mathbf{\hat{X}}, \mathbf{\hat{X}}} - \mathbf{K}_{\mathbf{\hat{X}}, \mathbf{X}}\mathbf{K}_{\mathbf{X}, \mathbf{X}}^{-1}\mathbf{K}_{\mathbf{X}, \mathbf{\hat{X}}}
	\label{eq:predictive-covariance}
\end{equation}
whose diagonal encodes prediction uncertainty.

\subsection{Cholesky Decomposition}
\label{subsec:cholesky-decomposition}
In Equations~\ref{eq:predictive-mean-simplified} and \ref{eq:predictive-covariance} for determining the predictive mean and covariance, respectively, $\mathbf{K}_{\mathbf{X}, \mathbf{X}}^{-1}$ occurs.
Computing inverse matrices explicitly is numerically unstable~\cite{Rasmussen2005_gp_ml} and has cubic complexity.

For instance, GPyTorch addresses this challenge with approximate methods, namely \gls{bbmm}~\cite{Gardner2018_gpytorch} for lower complexity inference, and LanczOs Variance Estimates~\cite{Pleiss2018_gpytorch_love} for the approximation of $\mathbf{\Sigma}$.
However, GPRat focuses on exact computations and therefore relies on the Cholesky decomposition to compute the mean and the posterior covariance matrix, still at cubic complexity but in a more numerically stable manner than classical inversion.
GPyTorch and GPflow also offer the Cholesky decomposition for exact computations.
In the benchmarks conducted in this work, both are configured to calculate predictions and uncertainties via Cholesky decomposition.

Cholesky decomposition can be applied to the symmetric and positive definite $\mathbf{K}_{\mathbf{X}, \mathbf{X}}$, and describes it as
\begin{equation}
	\mathbf{K}_{\mathbf{X}, \mathbf{X}} = \mathbf{L}\mathbf{L}^T
\end{equation}
where $\mathbf{L}$ is a lower triangular matrix, the so-called \emph{Cholesky factor}.
We first consider Equation~\ref{eq:predictive-mean-simplified} for the predictive mean.
Solving
\begin{equation}
	\mathbf{L} \vec{\beta} = \vec{y}
\end{equation}
and then
\begin{equation}
	\mathbf{L}^T \vec{\alpha} = \vec{\beta}
\end{equation}
allows computing $\mathbf{K}_{\mathbf{X}, \mathbf{X}}^{-1}\vec{y}$.
Furthermore, with
\begin{equation}
	\mathbf{L} \mathbf{V} = \mathbf{K}_{\mathbf{X}, \mathbf{\hat{X}}}
	\label{eq:l}
\end{equation}
and
\begin{equation}
	\mathbf{W} = \mathbf{V}^T\mathbf{V},
	\label{eq:w-subst}
\end{equation}
the posterior covariance matrix can be computed via
\begin{equation}
	\Sigma = \mathbf{K}_{\mathbf{\hat{X}}, \mathbf{\hat{X}}} - \mathbf{W}.
	\label{eq:sigma}
\end{equation}

\subsection{Tiled Algorithms}
\label{subsec:tiled-algorithms}

GPyTorch and GPflow compute the Cholesky decomposition monolithically, i.e., on a single tile and using their \gls{blas} backends for parallelization.
In contrast, GPRat is based on a tiled Cholesky algorithm.
In the following, the resulting tiled algorithm is sketched.
While the other algorithms required for computing Equations~\ref{eq:predictive-covariance} and \ref{eq:predictive-mean-simplified} also become tiled, only the Cholesky decomposition is considered here, as it is the central component.
GPRat utilizes a highly parallel right-looking strategy~\cite{Buttari2009_parallel_exakt_solvers}.

For brevity, we declare $\mathbf{K} = \mathbf{K}_{\mathbf{X}, \mathbf{X}}$.
The covariance matrix is decomposed into $M \times M$ tiles $\mathbf{K}_{I,J} \in \mathbb{R}^{m \times m}$ with $m=N/M$, and $I,J=0,\dots,M-1$.
Since the matrix is symmetric, it suffices to store the lower half of the tiles, including the main diagonal.
This way, memory savings of at least \qty{25}{\percent} for two tiles per dimension and up to \qty{50}{\percent} in the limit are possible in comparison to storing the entire matrix~\cite{Moellmann2026_gprat_gpu}.
Similarly, the Cholesky factor is a lower-triangular matrix, computed in place to save memory.

The algorithm applies standard level-3 \gls{blas} operations to these tiles, leveraging optimized memory access patterns~\cite{portf-stuff-optimized-for-memacc}.
In particular, \textcolor{potrf}{POTRF} (\textcolor{potrf}{PO}sitive Definite \textcolor{potrf}{TR}iangular \textcolor{potrf}{F}actorization), a \gls{lapack} operation commonly realized by \gls{blas} calls itself, \textcolor{trsm}{TRSM} (\emph{\textcolor{trsm}{Tr}iangular \textcolor{trsm}{S}olve with \textcolor{trsm}{M}ultiple Right-Hand Sides}), \textcolor{syrk}{SYRK} (\emph{\textcolor{syrk}{Sy}mmetric \textcolor{syrk}{R}ank-\textcolor{syrk}{K} Update}), and \textcolor{gemm}{GEMM} (\emph{\textcolor{gemm}{Ge}neral \textcolor{gemm}{M}atrix-\textcolor{gemm}{M}atrix Multiplication}) are used for the right-looking Cholesky decomposition.
\textcolor{gemm}{GEMM} calls dominate this workload: for the right-looking algorithm with $M$ tiles per dimension (Figure~\ref{fig:tiled-algo}), the number of \textcolor{gemm}{GEMM} calls grows as $\mathcal{O}(M^3)$, compared to $\mathcal{O}(M)$ for \textcolor{potrf}{POTRF} and $\mathcal{O}(M^2)$ for \textcolor{trsm}{TRSM} and \textcolor{syrk}{SYRK}.

A pseudocode for the tiled Cholesky decomposition is shown in Figure~\ref{fig:tiled-algo}, and a visualization of the application on the tiles is illustrated in Figure~\ref{fig:tiled-vis}.
The other tiled algorithms in GPRat also use level-2 and level-1 \gls{blas} operations and, where needed, custom CPU, CUDA, or SYCL kernels.

\section{SYCL Backend Implementation}
\label{sec:implementation}

This section describes how the SYCL backend implements the prediction methods, integrates oneMath, and manages data dependencies.

\subsection{GPRat Prediction Methods}
\label{subsec:gprat-prediction-methods}
All GPRat backends expose three different prediction methods, namely \texttt{predict}, \texttt{predict\allowbreak\_with\allowbreak\_uncertainty}, and \texttt{predict\allowbreak\_with\allowbreak\_full\_cov} in both Python and C\texttt{++}.
The former returns only the prediction, whereas the latter two propose different methods for computing uncertainty.
While both are mathematically exact, choosing \texttt{predict\allowbreak\_with\allowbreak\_uncertainty} over \texttt{predict\allowbreak\_with\allowbreak\_full\_cov} can yield a significant performance improvement.
We quantify this improvement as a speedup
\begin{equation}\label{eq:speedup}
	S= \frac{T_{\text{full\_cov}}} { T_{\text{uncertainty}}},
\end{equation}
where $T_{\text{full\_cov}}$ and $T_{\text{uncertainty}}$ denote the runtimes of \texttt{predict\allowbreak\_with\allowbreak\_full\_cov} and \texttt{predict\allowbreak\_with\allowbreak\_uncertainty}, respectively.
Thus, $S>1$ indicates a performance improvement from using \texttt{predict\allowbreak\_with\allowbreak\_uncertainty}, and $S<1$ a deterioration.

The difference lies in which parts of the posterior covariance matrix are computed.
\texttt{predict\allowbreak\_with\allowbreak\_full\_cov} computes the entire posterior covariance matrix.
This is equivalent to full-covariance Cholesky-based predictions in GPyTorch and GPflow.

On the other hand, tiled approaches, as implemented in GPRat, are more flexible in the computations they perform.
For prediction uncertainty, only the main diagonal of the posterior covariance matrix is relevant, so it suffices to compute only the corresponding tiles.
Computing the full covariance matrix is only beneficial if it is explicitly used for further processing.
As explicit use of the full covariance matrix is rarely the target application, we treat \texttt{predict\allowbreak\_with\allowbreak\_uncertainty}, which only computes the diagonal tiles, as the direct competitor to the diagonal-only Cholesky-based predictions in GPyTorch and GPflow in this work.

\subsection{Integration of oneMath Into GPRat}

GPRat is integrated into both Python and C\texttt{++} projects by creating a \texttt{GP} object exposing the methods introduced in Section~\ref{subsec:gprat-prediction-methods}.
The SYCL backend relies on the same \texttt{GP} constructor as the CUDA backend.
The two backends are mutually exclusive, and which one is used must be determined at compile time.
Internally, the underlying implementation of the exposed methods is chosen in the following way:
\begin{itemize}
	\item With \texttt{GPRAT\_WITH\_CUDA} or \texttt{GPRAT\_WITH\_SYCL}, the user can decide whether GPRat is compiled with CUDA or SYCL support.
	\item The CPU is always an available target, regardless of which of the aforementioned macros are defined.
	The target to use can be set at runtime using the corresponding constructor.
\end{itemize}
If the SYCL backend is used, the GPU vendor must be specified with an additional CMake option to determine which libraries to link against.
For AMD GPUs, the HIP target must be set, e.g., \texttt{"gfx90a"} for the AMD Instinct MI210 used in this work.

In general, the SYCL backend is structurally very similar to the CUDA backend.
However, it differs in how it manages data dependencies.
This impacts the internal structure and the method signatures.
In addition, although equivalent in content, the SYCL implementations of the custom kernels in GPRat operate on one-dimensional \texttt{sycl::range}s, which require neither an explicit block-size specification nor padding.

Regardless of whether CUDA or SYCL support is enabled, both the oneMKL and the more portable OpenBLAS backend for the CPU variant can be chosen.
Since the OpenBLAS backend already ensures portability on CPUs, we consider the SYCL backend to be a GPU backend first and foremost.
The compile infrastructure includes NVIDIA, AMD, and Intel GPUs, all three of which were explicitly tested in this work.

\subsection{Data Dependency Management}
\label{subsec:data-dependency-management}
Both SYCL and CUDA offer multiple scheduling strategies.
With CUDA, algebraic kernels are scheduled through streams.
Multiple streams can target the same GPU, and each stream targets exactly one GPU.
All kernels submitted to a stream are executed in the order they are enqueued.
In contrast, kernels submitted to different streams are unordered with respect to one another and may execute concurrently, subject to available device resources.
CUDA streams schedule kernel launches only on target devices and do not handle data dependencies.
CUDA addresses this shortcoming through events~\cite{nvidia2025_cuda}, which allow kernels to be scheduled explicitly.
With CUDA 10.0~\cite{cuda-graphs-in-cuda-10}, CUDA graphs~\cite{nvidia2025_cuda} were introduced, allowing dependencies to be declared once as an asynchronous task graph and replayed with a single launch.
In contrast, GPRat's CUDA backend takes another approach.
The backend instead manages dependencies using HPX's \texttt{hpx::dataflow} functionality~\cite{Moellmann2026_gprat_gpu} to incorporate the kernels directly into the HPX task graph.

Wrapping single oneMath \textcolor{gemm}{GEMM} calls in \texttt{hpx::dataflow} or \texttt{hpx::async} produced segmentation faults.
In contrast, the same pattern works reliably for GPRat's custom SYCL kernels, suggesting the conflict lies specifically in oneMath's backend libraries rather than in HPX/SYCL interoperability in general.
As a workaround, we run the oneMath calls outside the HPX task graph.
Each oneMath wrapper draws a queue round-robin from a shared pool, submits its operation, and calls \texttt{queue.wait()} before submitting the next tile's operation.
HPX and SYCL interoperate at the task granularity here, not through a shared event model: each \texttt{hpx::async} task wraps a device operation and completes once that operation finishes on its worker thread, which is sufficient to achieve the intended per-tile overlap across the pipeline's independent stages.
HPX dispatches one task per tile, orchestrating both tile assembly and the custom kernels used for uncertainty estimation.

\section{Results}
\label{sec:results}
This section analyzes the performance of various setups across problem-size scaling scenarios.
First, we compare the GPRat-internal \texttt{predict\allowbreak\_with\allowbreak\_uncertainty} and \texttt{predict\allowbreak\_with\allowbreak\_full\_cov} functions to quantify the potential performance improvement.
Then, we compare the GPRat backends against each other and against reference GPflow and GPyTorch implementations, and finally analyze the underlying Cholesky kernel in isolation.
The tests were executed on an AMD EPYC 9274F dual-socket system, an NVIDIA A30, an AMD Instinct MI210, and an Intel Arc B580.
Thorough information on the systems and software versions used is provided in Appendix~\ref{sec:reproducibility}.
In addition to the target hardware, a setup consists of a core count of 48 for the CPU backend, a problem size $N = \hat{N} = 2^i, i = 0,1,\dots,14$, and, in the case of GPRat, several tile counts per dimension in $\{1,2,4,8,16,32\}$.
GPU-based configurations, including the GPflow and GPyTorch reference implementations, use a single CPU core for host-side orchestration as compute is handled on the device.

Since we focus on the predictions, hyperparameter optimization was not applied, and all scenarios were tested with eight regressors using the provided \gls{msd} dataset~\cite{og-gprat-data-paper}, i.e., a system of nonlinearly coupled carts whose lagged states serve as regressors, following a Nonlinear Finite-Impulse Response model~\cite{Kocijan2016_gp_si_book}.
Both GPyTorch and GPflow are configured to use Cholesky decomposition for predictions, which ensures a fair algorithmic comparison with GPRat.
Note that this disables GPyTorch's approximate default \gls{bbmm} inference mode.
Thus, the results presented in this work do not reflect GPyTorch's default performance.
We tested each configuration with ten repeated runs, and the plots show the median runtime with min-max error bars.
All tests were carried out with double precision.

GPRat's CPU backend uses oneMKL, whereas on the A30, both the CUDA and SYCL backends are tested.
The MI210 and the Arc B580 use the SYCL backend.
The CUDA backend was configured with a single stream.
Likewise, the SYCL backend operates on a single queue.

\subsection{Uncertainty Versus Full Covariance Matrix}
\label{subsec:uncer-vs-full}
The speedup $S$ (Equation \ref{eq:speedup}) achieved by prediction with uncertainty over the full-covariance approach varies by device and problem size.
We report the median of ten repeated runs, with the min-max range in parentheses.

For CUDA on the A30, median speedups remain close to one for small problem sizes.
With the SYCL backend on the A30, the median speedup is similarly close to 1 for small problem sizes, but we observed a much wider runtime range than the CUDA backend on the same hardware.

For small problem sizes, the uncertainty-only path skips certain \textcolor{gemm}{GEMM} calls but still incurs the fixed launch overhead for the remaining kernels, resulting in a smaller saving than the added cost.
At larger scales, the compute savings dominate, and oneMath/SYCL overheads become irrelevant.
On the Intel Arc B580, the median speedup is between $0.70$ and $0.85$ across nearly the entire small-to-medium range.
The median speedup exceeds 1 only from problem size \num{1024} onward, later than on any other tested device.

For problem sizes \num{2048} and above, GPU median speedups increase steadily.
At the largest evaluated size they range from $1.42$ (range $1.40$--$1.45$) on the MI210 with oneMath to $2.39$ (range $2.36$--$2.40$) on the A30 with CUDA, with the Intel Arc B580 reaching $2.38$ (range $2.36$--$2.40$).
Meanwhile, on the CPU, the median speedup ranges from $2.2$ to $2.3$ from problem size \num{8192} onward.
With a median speedup of at least $1.42$ for the maximum problem size without losing accuracy, we consider this simplification successful.
These results are expected given how the simplifications introduced by predictive uncertainty come into effect.
In brief, Equations~\ref{eq:w-subst} and \ref{eq:sigma} are only solved on tiles containing elements of the main diagonal.
This way, the uncertainty-only path saves many \textcolor{gemm}{GEMM} calls, depending on both the training and the test size.
In turn, the benefit for small problem sizes is limited.
As the problem size increases, the off-diagonal tiles grow, so skipping them yields a larger speedup.
Consequently, the speedup from prediction with uncertainty increases with problem size, reaching up to $2.39$ at the largest sizes evaluated.

\subsection{Prediction with Uncertainty}
\label{subsec:pred-uncertainty-results}
In the following, the prediction runtimes with uncertainty are first compared across different GPRat setups, and second, the runtimes of GPRat and its competitors on GPUs are surveyed.

\subsubsection{GPRat Evaluation}
\begin{figure*}[t]
	\centering
	\begin{minipage}{0.48\textwidth}
		\includegraphics[width=\linewidth]{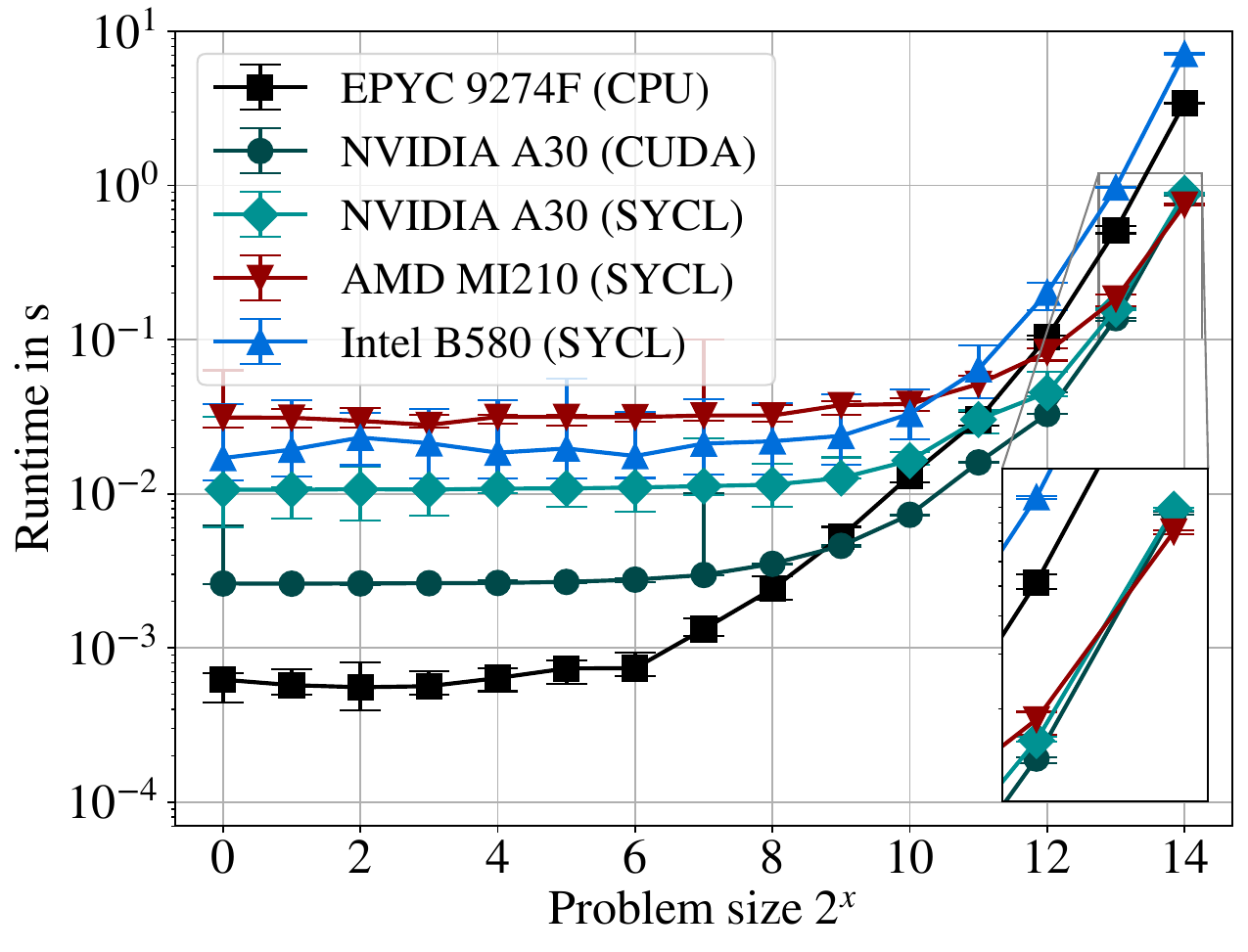}
		\caption{GPRat problem size scaling for prediction with uncertainty on all tested devices, optimal configuration per problem size.}
		\label{fig:pred_uncer_all}
	\end{minipage}
	\hfill
	\begin{minipage}{0.48\textwidth}
		\includegraphics[width=\linewidth]{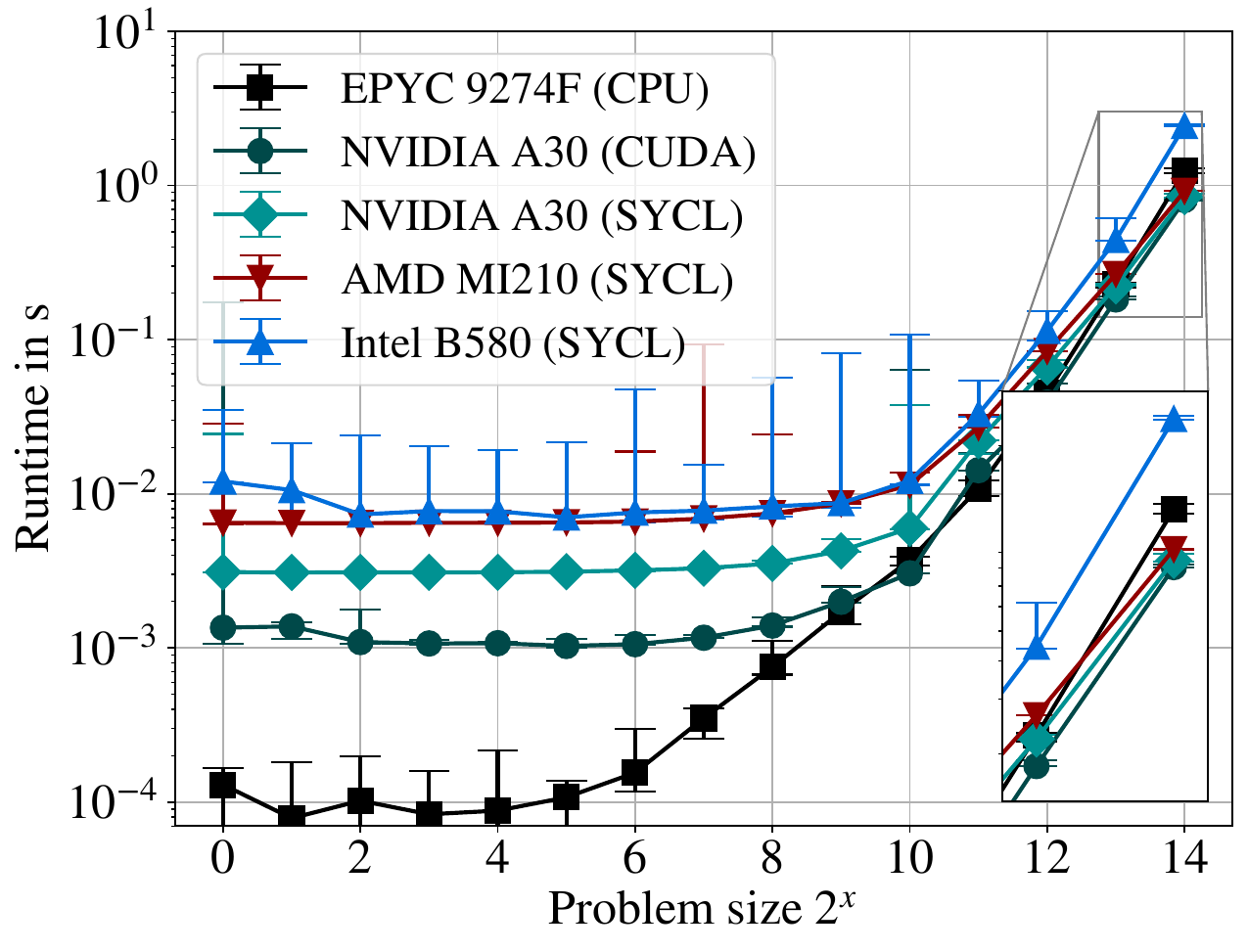}
		\caption{GPRat problem size scaling for Cholesky decomposition on all tested devices, optimal configuration per problem size.}
		\label{fig:cholesky_all}
	\end{minipage}
\end{figure*}

The runtimes of prediction with uncertainty with GPRat on all tested devices are depicted in Figure~\ref{fig:pred_uncer_all}.

\begin{figure*}[t]
	\begin{minipage}{0.48\textwidth}
		\includegraphics[width=\linewidth]{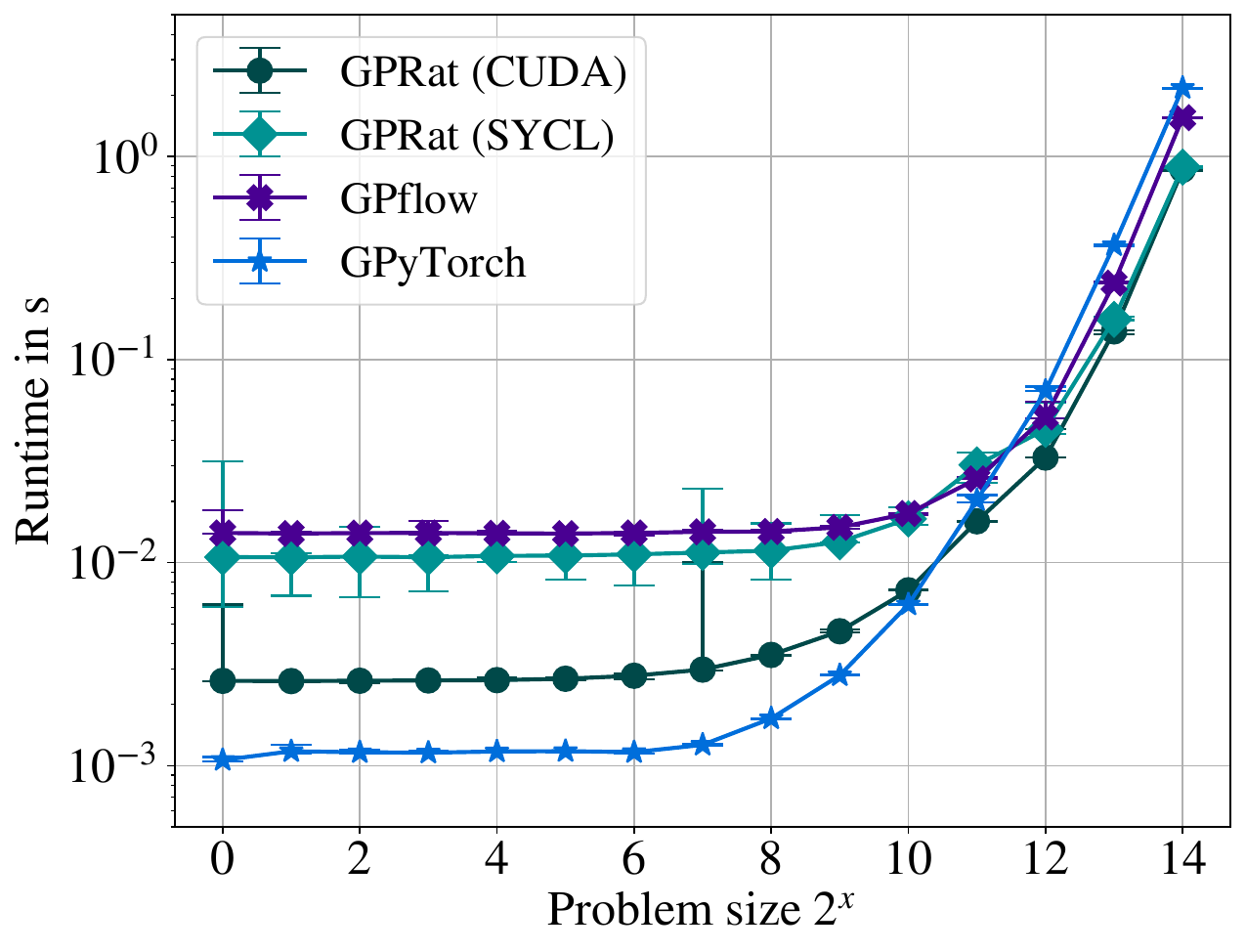}
		\caption{Problem size scaling comparison for prediction with uncertainty on the NVIDIA A30, optimal configuration per problem size.}
		\label{fig:pred_uncer_a30}
	\end{minipage}
	\hfill
	\begin{minipage}{0.48\textwidth}
		\includegraphics[width=\linewidth]{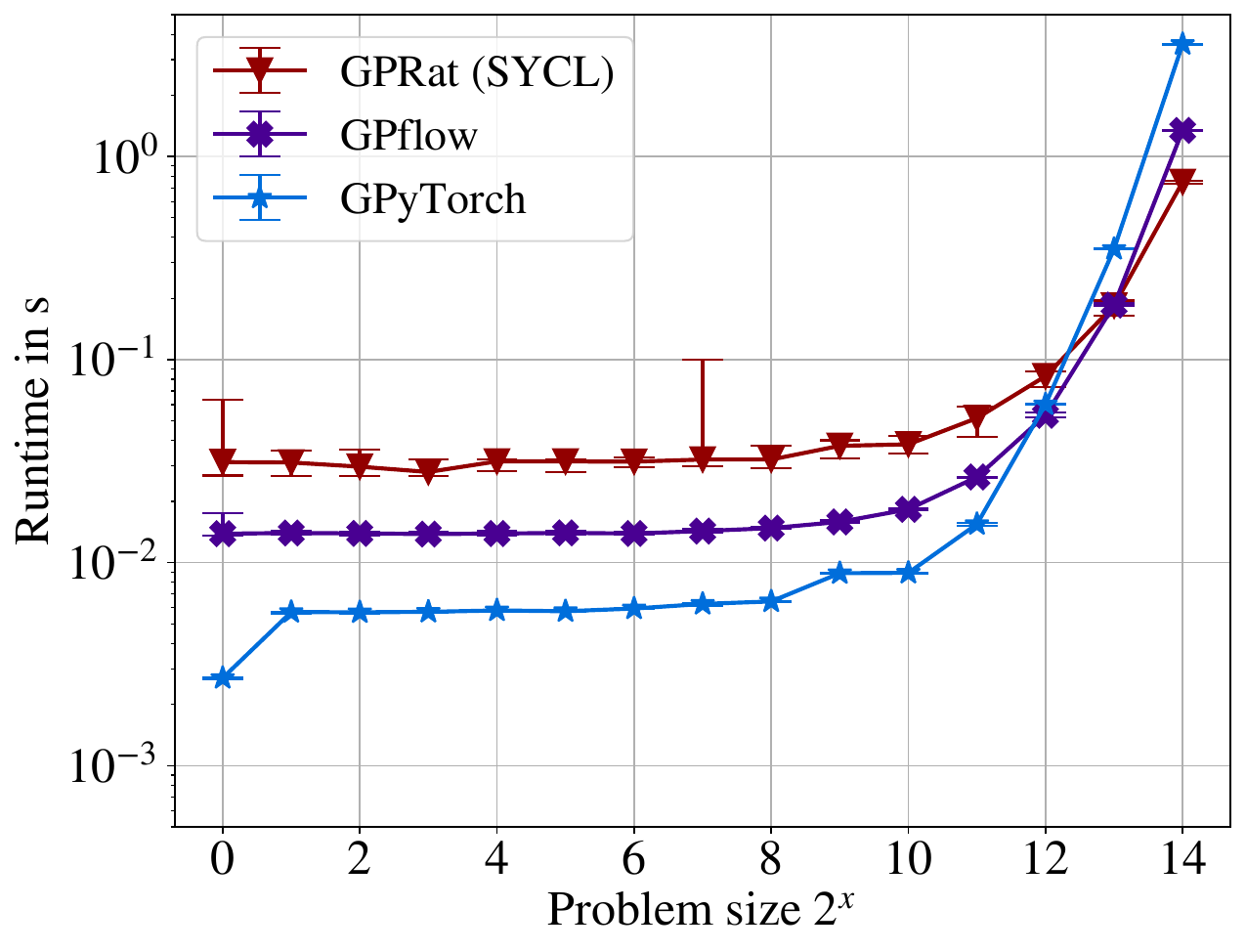}
		\caption{Problem size scaling comparison for prediction with uncertainty on the AMD Instinct MI210, optimal configuration per problem size.}
		\label{fig:pred_uncer_mi210}
	\end{minipage}
\end{figure*}

\paragraph{CPU (oneMKL backend)}
The CPU results use \num{48} physical cores of the reference system, with HPX worker threads pinned one-to-one to physical cores and oneMKL configured to run each \gls{blas} call sequentially, so that parallelism is provided entirely by GPRat's own tile-level task scheduling rather than by nested threading inside oneMKL.
For problem sizes up to \num{512}, the CPU achieves the lowest median runtimes of \qtyrange{0.24}{3.51}{\milli\second}, outperforming all GPU setups.
This can be explained by memory allocation overheads, which dominate execution time for small problem sizes.
Each allocation synchronizes the device, and tiles are allocated sequentially.
In this case, prediction can be scaled downward more efficiently on the CPU, where \emph{downward scalability} refers to the ability to handle small problem sizes with low overhead.
For problem sizes \num{64} to \num{2048}, scaling is nearly ideal, but it slightly degrades at larger sizes.
At this point, the inferior theoretical peak floating-point performance, the limited memory bandwidth, and the lower level of CPU parallelism become apparent.
For the largest tested problem size of \num{16384}, the median runtime is \qty{3.43}{\second}.

\paragraph{NVIDIA GPU (CUDA and SYCL backends)}
As expected, the native CUDA backend is the more performant on the A30.
Until problem size \num{8192}, it is either the most or second-most performant GPRat backend.
For small problem sizes, we observe two patterns.
First, the SYCL backend has noticeably wider error bars than the CUDA backend.
Second, the runtimes of both backends have a plateau, reflecting the aforementioned limited downward scalability of GPRat's GPU implementation.
At scale, the relative overhead of the SYCL backend becomes negligible, and the runtime gradient is similar to that of the CPU.
Nevertheless, for the largest problem size, the A30 is much faster than the CPU, requiring \qty{861}{\milli\second} and \qty{887}{\milli\second} with CUDA and oneMath, respectively.
For the largest tested problem size ($n=\num{16384}$), the SYCL backend requires only approximately \qty{3}{\percent} more time than the CUDA backend.
Compared to the \qty{3.43}{\second} of the CPU, this results in speedup factors of $3.98$ (CUDA) and $3.87$ (oneMath).
Notice that the speedup exceeds what the higher peak performance would indicate, revealing scaling issues in the CPU version.

\paragraph{AMD (SYCL backend)}
An initial runtime plateau can be observed, ranging from \qtyrange{28.0}{37.6}{\milli\second}, which is higher than on the A30.
While the MI210 has larger error bars than the A30 with CUDA, it has significantly smaller ones than the A30 with oneMath.
The SYCL backend generally increases runtime variance, with the effects becoming especially apparent on the A30.
Isolating the residual difference between the A30 and the MI210 in detail would require device-level queue tracing, which we leave for future work.

Although the MI210 is the slowest option for small problem sizes, it achieves the lowest runtime for the largest problem size, at \qty{751}{\milli\second}, yielding a speedup of $1.18$ over the A30's SYCL backend.
Considering that the MI210's peak FP64 performance is roughly four times higher, these results indicate inefficient resource utilization.
The speedup factor over the CPU is $4.57$.

\paragraph{Intel GPU (SYCL backend)}
On the Arc B580, GPRat exhibits the widest dynamic range of any GPU backend tested, but its behavior at scale is directly tied to its hardware class.
Relative to the reference CPU, the Arc B580 has only $0.52\times$ the FP64 throughput.
Relative to the A30 and MI210, it has $0.33\times$ and $0.076\times$ their FP64 peaks.
For small problem sizes, an initial runtime plateau of \qtyrange{17}{24}{\milli\second} is observed, below the MI210's plateau (\qtyrange{28.0}{37.6}{\milli\second}).
As the problem size grows, the hardware deficit takes over: at the largest tested problem size, the Arc B580 requires \qty{7.14}{\second}, the slowest GPU configuration evaluated, a factor of $9.51$ slower than the fastest GPU (the MI210, at \qty{751}{\milli\second}) and, notably, a factor of about $2.08$ slower than GPRat's own CPU backend (\qty{3.43}{\second}).
This is the only GPU configuration in this work that does not outperform the CPU at the largest evaluated problem size, and the fact that its FP64 peak trails the reference CPU's makes that outcome expected rather than anomalous.

\subsubsection{GPRat, GPflow, and GPyTorch Comparison}
All three implementations are configured for exact inference and compute mathematically equivalent predictions and uncertainties, so the runtime differences discussed below reflect implementation choices rather than differences in the underlying computation.
GPRat's reference benchmarking scripts for GPflow and GPyTorch, available alongside the reproducibility artifacts (Appendix~\ref{sec:reproducibility}), turn off GPyTorch's default approximate-inference path (\gls{bbmm}) so that all three compute the same Cholesky-based quantities.
GPflow's \texttt{GPR} model is exact by default and requires no equivalent change.
This configuration does not make the diagonal-only path equally cheap for both libraries.
Specifically, GPflow's diagonal-only predictions (\texttt{full\_cov=False}) genuinely avoid forming the full posterior covariance matrix. In contrast, GPyTorch's diagonal-only prediction internally constructs a dense covariance-correction term comparable to its full-covariance prediction, analogous to $\mathbf{W}$ in Equation~\ref{eq:w-subst}, unless lazy kernel evaluation is left enabled.
In our testing, the lazy configuration is $1.13\times$ faster than the non-lazy configuration on the A30 and $1.03\times$ faster on the MI210.
For comparison, GPflow's diagonal-only path is $1.82\times$ (A30) and $1.27\times$ (MI210) faster than its own full-covariance computation at the same problem size.
Thus, we present results for lazy-evaluated GPyTorch and diagonal-only GPflow.
The following comparison should therefore be read as a controlled comparison between specific configurations under matching exact-inference settings, not as a general verdict on the relative performance of GPRat, GPflow, and GPyTorch as libraries.
A comparison between GPRat, GPflow, and GPyTorch on the A30 and the MI210 is shown in Figures~\ref{fig:pred_uncer_a30} and \ref{fig:pred_uncer_mi210}, respectively.
This comparison excludes the Arc B580: At the time of writing, PyTorch's Intel XPU backend does not yet provide a GPU implementation of the Cholesky-based linear algebra operations that exact-inference GPyTorch requires, so GPyTorch's linear-algebra-heavy core falls back to the CPU rather than exercising the GPU.
A corresponding GPflow/TensorFlow comparison on the Arc B580 is currently not possible either.
Intel GPU execution is available only through the Intel Extension for TensorFlow~\cite{IntelTensorFlow}, which does not officially support the Arc B580 in its most recent release.

\paragraph{NVIDIA GPU}
For small problem sizes, GPyTorch and GPflow are the fastest and the slowest options, respectively.
The former is up to $13.2\times$ faster than the latter.
In turn, the runtimes of GPyTorch rise from an initial plateau and eventually approach those of GPflow.
At scale, GPflow outperforms GPyTorch with a runtime of \qty{1.54}{\second} vs.\ \qty{1.92}{\second}, resulting in a speedup of approximately $1.25$.
While GPyTorch allows better downward scalability, GPflow performs better at scale.
Meanwhile, the runtimes of GPRat remain between those of its competitors for small problem sizes, whereas at scale, they are significantly lower.
For the largest tested problem size, the SYCL backend achieves a runtime of \qty{887}{\milli\second}, offering speedups of $1.75$ over GPflow and $2.17$ over GPyTorch.
In conclusion, GPRat outperforms its main competitor, GPflow, at large problem sizes and can offer improvements for small ones as well.

\paragraph{AMD GPU}
On the MI210, GPflow and GPyTorch are closer in terms of performance than on the A30.
Again, the former is slower for small problem sizes but faster at scale.
The performance of GPflow on the MI210 is rather similar to that on the A30, with an initial plateau around \qty{13.4}{\milli\second} and a runtime of \qty{1.35}{\second} for the largest problem size.
For large problem sizes, GPyTorch exhibits a steep increase in runtime.
GPRat is the slowest option for small- to medium-sized problems but the fastest at scale, with the shallowest runtime gradient.
With \qty{751}{\milli\second}, the speedup over GPflow and GPyTorch is $1.79$ and $4.62$, respectively.
Except at a large scale, all implementations tend to perform worse than on the A30.
This puts the comparatively low performance of GPRat on the MI210 into perspective relative to the other devices shown in Figure~\ref{fig:pred_uncer_all}.

A native HIP/rocBLAS baseline on the MI210 was not measured.
However, oneMath on AMD hardware dispatches \gls{blas} operations to rocBLAS internally, analogous to its dispatch to cuBLAS on NVIDIA.
The SYCL-to-native overhead therefore resides primarily in the SYCL runtime dispatch layer rather than the underlying math library.
The MI210 overhead is therefore plausibly of comparable magnitude, though it was not directly measured, and this cannot be confirmed from the NVIDIA results given the different drivers and dispatch layer.

\subsection{Cholesky Decomposition}
\label{subsec:cholesky}
In the following, a problem-size-scaling benchmark of the tiled Cholesky decomposition with GPRat is presented.
The results are depicted in Figure~\ref{fig:cholesky_all}.
The performance of the devices for the Cholesky decomposition itself is compared to that for the entire prediction with uncertainty depicted in Figure~\ref{fig:pred_uncer_all}.
Once more, the CPU yields the best performance for small problem sizes, with a noticeable advantage over the GPU configurations.
However, execution on the A30 becomes the superior option earlier than for prediction with uncertainty.
For the largest tested problem size, the CPU requires \qty{1.24}{\second}, whereas the GPU runtimes range from \qty{0.81}{\second} on the A30 to \qty{0.92}{\second} on the MI210, i.e., speedups of $1.54$ and $1.35$, respectively.
The prediction with uncertainty allows for even higher relative speedups on the A30 and MI210.

The speedups reported differ significantly from those in~\cite{Moellmann2026_gprat_gpu}.
In the case of the latter, only the runtimes of the Cholesky decomposition itself were considered.
The time required to copy data to and from the host, especially to copy the Cholesky factor back to the host, was not accounted for, resulting in lower GPU runtimes.

For the Cholesky decomposition, the error bars of the SYCL backend are mostly narrow on both the MI210 and the A30.
This contrasts with the observations for the prediction with uncertainty, where the SYCL backend showed noticeable runtime variance, especially on the A30.
Therefore, additional runtime fluctuations originate from the other tiled algorithms involved in the prediction.

The Arc B580 is the slowest configuration for the Cholesky decomposition across the entire tested range, including relative to the CPU.
This is expected given the same FP64-throughput comparison established in Section~\ref{subsec:pred-uncertainty-results}, together with the memory-bandwidth figures in Table~\ref{table:reproducibility-hardware}: the Arc B580 trails the CPU on FP64 throughput, whereas the A30 and MI210 exceed the CPU on both FP64 throughput and memory bandwidth, comfortably so for the MI210 and by a slimmer margin for the A30's bandwidth.
At the largest problem size, it requires \qty{2.44}{\second}, a factor of $3.03$ slower than the fastest GPU (the A30 with CUDA, at \qty{806}{\milli\second}) and a factor of $1.97$ slower than the CPU.
We also encountered a related software-maturity issue during development: the first just-in-time compilation of a kernel on the Arc B580's Level-Zero driver could crash when it occurred amid concurrent kernel dispatch elsewhere in the backend, which we worked around with a one-time warm-up compile issued ahead of the concurrent dispatch, a step the NVIDIA and AMD backends do not require.

\subsection{Limitations}

Several constraints bound the scope of these results.
First, the evaluation covers only \gls{gpr} prediction and uncertainty computation, whereas training through hyperparameter optimization was not measured.
Second, each backend was evaluated on a single GPU per vendor.
Third, no native HIP/rocBLAS baseline was measured on the MI210, so the SYCL portability overhead is quantified only against native CUDA on the A30.
Fourth, we could not test on an Intel Data Center GPU, and no native Level-Zero/oneMKL baseline outside oneMath was measured.
However, this limitation is a design choice, as this work aims to provide a portable, single-source solution for GPRat rather than a collection of vendor-specific backends.
Fifth, all results reflect one snapshot of a fast-moving software stack (Table~\ref{table:reproducibility-software}).

\section{Conclusion and Outlook}
\label{sec:conclusion}

In this work, we propose a SYCL backend for the GPRat library.
The proposed SYCL backend delivers portability across all three major GPU vendors through the unified SYCL API while maintaining competitive performance.
Its strengths are most pronounced at scale: the NVIDIA and AMD GPUs then outperform the CPU by a wide margin.
Furthermore, the SYCL backend's portability overhead relative to native CUDA on the A30 shrinks to approximately \qty{3}{\percent} at the largest problem size evaluated.
All GPRat backends compute mathematically exact predictions with uncertainty while operating only on the relevant diagonal tiles of the covariance matrix.
Combined with GPRat's tiled parallelism, this enables the SYCL backend also to outperform the corresponding exact functions of the established \gls{gpr} libraries GPflow and GPyTorch on both the A30 and the MI210.
For small problem sizes, however, GPyTorch in particular is more performant, and GPRat's CPU implementation offers better downward scalability than any GPU backend.
The portability overhead is likewise more noticeable at small problem sizes, and on both the A30 and the MI210, the portable SYCL backend shows higher runtime variance than the other tested configurations.

The Arc B580 does not outperform GPRat's CPU backend at the largest evaluated problem size, trailing it by about $2\times$ for both the Cholesky decomposition and prediction with uncertainty.
This result is largely explained by the hardware class rather than by a lack of performance portability: the Arc B580 is a consumer part whose FP64 peak trails the reference CPU's, so for a double-precision workload, it has no headroom over the CPU to begin with.

For all tested \gls{gpr} implementations, achieving good performance on the MI210 is more challenging than on the A30.
Despite having the highest double-precision peak performance and memory bandwidth among the tested devices, the MI210's small caches may limit data reuse, which the computationally intensive level-3 \gls{blas} kernels rely on.

The tiled-algorithm approach sets GPRat apart from the tested alternatives: it enables the diagonal-only simplification for prediction, with uncertainty exploited throughout this work, and even small tile counts yield at least \qty{25}{\percent} memory savings.
This is particularly valuable on GPUs, where memory is scarce and can limit the feasibility of certain computations.

Beyond these device- and algorithm-specific observations, the central insight of this work concerns the interaction between task-based runtimes and GPU APIs.
Wrapping single-tile oneMath \gls{blas} calls directly in HPX tasks proved unstable.
Keeping oneMath \gls{blas} calls outside the HPX task graph was the decisive choice behind this hybrid HPX/SYCL design.
This design choice enables single-source portability across NVIDIA, AMD, and Intel GPUs.

Several directions remain for future work.
First, and most importantly, the benchmarks here cover only \gls{gpr} prediction.
Evaluating the SYCL backend under offline hyperparameter optimization (training) is the key next step toward a complete performance picture.
Second, evaluating the backend on officially supported Intel hardware, such as the Intel Data Center GPU Max series, is a logical next step, ideally combined with device-level profiling.
Third, extending GPRat to distributed systems would enable operation on heterogeneous multi-node clusters with GPUs from multiple vendors.
Fourth, repeating this evaluation on a newer software stack including ROCm 7.x,  recent \texttt{icpx} releases, and ideally newer CDNA hardware such as the MI300 would separate hardware from software effects in the observed AMD behavior.

\section*{AI Usage Disclosure}
Generative AI tools, including Grammarly~\cite{grammarly} and Claude~\cite{claude}, were employed to enhance the clarity, grammar, and overall coherence of the manuscript.
All technical content, data analyses, and research findings were conceived and developed independently by the authors.
AI-assisted outputs were carefully reviewed, verified, and edited by the authors to ensure factual accuracy, interpretive rigor, and scholarly integrity.
The final manuscript reflects the authors' original intellectual contributions and analytical work.

\bibliographystyle{IEEEtran}
\bibliography{bibliography}

\begin{thebibliography}{10}
\providecommand{\url}[1]{#1}
\csname url@samestyle\endcsname
\providecommand{\newblock}{\relax}
\providecommand{\bibinfo}[2]{#2}
\providecommand{\BIBentrySTDinterwordspacing}{\spaceskip=0pt\relax}
\providecommand{\BIBentryALTinterwordstretchfactor}{4}
\providecommand{\BIBentryALTinterwordspacing}{\spaceskip=\fontdimen2\font plus
\BIBentryALTinterwordstretchfactor\fontdimen3\font minus
  \fontdimen4\font\relax}
\providecommand{\BIBforeignlanguage}[2]{{%
\expandafter\ifx\csname l@#1\endcsname\relax
\typeout{** WARNING: IEEEtran.bst: No hyphenation pattern has been}%
\typeout{** loaded for the language `#1'. Using the pattern for}%
\typeout{** the default language instead.}%
\else
\language=\csname l@#1\endcsname
\fi
#2}}
\providecommand{\BIBdecl}{\relax}
\BIBdecl

\bibitem{top500}
{TOP500.org}. (2026) {TOP500 List, June 2026}. [Online]. Available:
  \url{https://top500.org/lists/top500/2026/06/}. [Accessed: Aug. 5, 2026].

\bibitem{SyclSpec}
\BIBentryALTinterwordspacing
M.~Rovatsou, L.~Howes, R.~Keryell \emph{et~al.} (2025) {SYCL™ 2020
  Specification}. [Online]. Available:
  \url{https://registry.khronos.org/SYCL/specs/sycl-2020/html/sycl-2020.html}
\BIBentrySTDinterwordspacing

\bibitem{OneMathSpec}
\BIBentryALTinterwordspacing
{UXL Foundation}. (2026) {oneMath specification}. [Online]. Available:
  \url{https://oneapi-spec.uxlfoundation.org/specifications/oneapi/latest/elements/onemath/source/}
\BIBentrySTDinterwordspacing

\bibitem{Rasmussen2005_gp_ml}
\BIBentryALTinterwordspacing
C.~E. Rasmussen and C.~K.~I. Williams, \emph{Gaussian Processes for Machine
  Learning}.\hskip 1em plus 0.5em minus 0.4em\relax The MIT Press, 11 2005.
  [Online]. Available: \url{https://doi.org/10.7551/mitpress/3206.001.0001}
\BIBentrySTDinterwordspacing

\bibitem{Kocijan2016_gp_si_book}
J.~Kocijan, \emph{Modelling and Control of Dynamic Systems Using Gaussian
  Process Models}, ser. Advances in Industrial Control.\hskip 1em plus 0.5em
  minus 0.4em\relax Springer International Publishing, 01 2016.

\bibitem{Mackay1998_intro_gp}
\BIBentryALTinterwordspacing
D.~J.~C. MacKay, ``{Introduction to Gaussian Processes},'' \emph{NATO ASI
  Series F: Computer and Systems Sciences}, vol. 168, pp. 133--166, 12 1998.
  [Online]. Available: \url{https://www.inference.org.uk/mackay/gpB.ps.gz}
\BIBentrySTDinterwordspacing

\bibitem{Williams1995_gpr}
C.~K.~I. Williams and C.~E. Rasmussen, ``Gaussian processes for regression,''
  in \emph{Proceedings of the 9th International Conference on Neural
  Information Processing Systems}, ser. NIPS'95.\hskip 1em plus 0.5em minus
  0.4em\relax Cambridge, MA, USA: MIT Press, 1995, p. 514–520.

\bibitem{Damianou2013_deep_gps}
\BIBentryALTinterwordspacing
A.~Damianou and N.~D. Lawrence, ``{Deep {G}aussian Processes},'' in
  \emph{Proceedings of the Sixteenth International Conference on Artificial
  Intelligence and Statistics}, ser. Proceedings of Machine Learning Research,
  C.~M. Carvalho and P.~Ravikumar, Eds., vol.~31.\hskip 1em plus 0.5em minus
  0.4em\relax Scottsdale, Arizona, USA: PMLR, 29 Apr--01 May 2013, pp.
  207--215. [Online]. Available:
  \url{https://proceedings.mlr.press/v31/damianou13a.html}
\BIBentrySTDinterwordspacing

\bibitem{Kennedy2001_gp_sc}
M.~C. Kennedy and A.~O'Hagan, ``{Bayesian Calibration of Computer Models},''
  \emph{Journal of the Royal Statistical Society Series B}, vol.~63, pp.
  425--464, 09 2001.

\bibitem{Liu2020_gp_bigdata}
H.~Liu, Y.-S. Ong, X.~Shen, and J.~Cai, ``{When Gaussian Process Meets Big
  Data: A Review of Scalable GPs},'' \emph{IEEE Transactions on Neural Networks
  and Learning Systems}, vol.~31, no.~11, pp. 4405--4423, 2020.

\bibitem{Deringer2021_gpr_chemistry}
\BIBentryALTinterwordspacing
V.~L. Deringer, A.~P. Bartók, N.~Bernstein, D.~M. Wilkins, M.~Ceriotti, and
  G.~Csányi, ``Gaussian process regression for materials and molecules,''
  \emph{Chemical Reviews}, vol. 121, no.~16, pp. 10\,073--10\,141, 2021, pMID:
  34398616. [Online]. Available:
  \url{https://doi.org/10.1021/acs.chemrev.1c00022}
\BIBentrySTDinterwordspacing

\bibitem{Schulz2018_gpr_tutorial}
\BIBentryALTinterwordspacing
E.~Schulz, M.~Speekenbrink, and A.~Krause, ``A tutorial on gaussian process
  regression: Modelling, exploring, and exploiting functions,'' \emph{Journal
  of Mathematical Psychology}, vol.~85, pp. 1--16, 2018. [Online]. Available:
  \url{https://www.sciencedirect.com/science/article/pii/S0022249617302158}
\BIBentrySTDinterwordspacing

\bibitem{Castro2023_python_slow}
O.~Castro, P.~Bruneau, J.-S. Sottet, and D.~Torregrossa, ``{Landscape of
  High-Performance Python to Develop Data Science and Machine Learning
  Applications},'' \emph{ACM Comput. Surv.}, vol.~56, no.~3, 10 2023.

\bibitem{Paszke2019_pytorch}
A.~Paszke, S.~Gross, F.~Massa, A.~Lerer, J.~Bradbury, G.~Chanan, T.~Killeen,
  Z.~Lin, N.~Gimelshein, L.~Antiga, A.~Desmaison, A.~K\"{o}pf, E.~Yang,
  Z.~DeVito, M.~Raison, A.~Tejani, S.~Chilamkurthy, B.~Steiner, L.~Fang,
  J.~Bai, and S.~Chintala, \emph{PyTorch: an imperative style, high-performance
  deep learning library}.\hskip 1em plus 0.5em minus 0.4em\relax Red Hook, NY,
  USA: Curran Associates Inc., 2019.

\bibitem{Abadi2016_tensorflow}
M.~Abadi, P.~Barham, J.~Chen, Z.~Chen, A.~Davis, J.~Dean, M.~Devin,
  S.~Ghemawat, G.~Irving, M.~Isard, M.~Kudlur, J.~Levenberg, R.~Monga,
  S.~Moore, D.~G. Murray, B.~Steiner, P.~Tucker, V.~Vasudevan, P.~Warden,
  M.~Wicke, Y.~Yu, and X.~Zheng, ``Tensorflow: a system for large-scale machine
  learning,'' in \emph{Proceedings of the 12th USENIX Conference on Operating
  Systems Design and Implementation}, ser. OSDI'16.\hskip 1em plus 0.5em minus
  0.4em\relax USA: USENIX Association, 2016, pp. 265--283.

\bibitem{Gardner2018_gpytorch}
J.~R. Gardner, G.~Pleiss, D.~Bindel, K.~Q. Weinberger, and A.~G. Wilson,
  ``Gpytorch: blackbox matrix-matrix gaussian process inference with gpu
  acceleration,'' in \emph{Proceedings of the 32nd International Conference on
  Neural Information Processing Systems}, ser. NIPS'18.\hskip 1em plus 0.5em
  minus 0.4em\relax Red Hook, NY, USA: Curran Associates Inc., 2018, p.
  7587–7597.

\bibitem{Matthews2017_gpflow}
\BIBentryALTinterwordspacing
A.~G. de~G.~Matthews, M.~van~der Wilk, T.~Nickson, K.~Fujii, A.~Boukouvalas,
  P.~Le{\'o}n-Villagr{\'a}, Z.~Ghahramani, and J.~Hensman, ``{GPflow: A
  Gaussian Process Library using TensorFlow},'' \emph{Journal of Machine
  Learning Research}, vol.~18, no.~40, pp. 1--6, 2017. [Online]. Available:
  \url{http://jmlr.org/papers/v18/16-537.html}
\BIBentrySTDinterwordspacing

\bibitem{Helmann2026_gprat}
\BIBentryALTinterwordspacing
M.~Helmann, A.~Strack, and D.~Pfl{\"u}ger, ``{GPRat: Gaussian Process
  Regression with Asynchronous Tasks},'' in \emph{Asynchronous Many-Task
  Systems and Applications}, P.~Diehl, Q.~Cao, T.~Herault, and G.~Bosilca,
  Eds.\hskip 1em plus 0.5em minus 0.4em\relax Cham: Springer Nature
  Switzerland, 2026, pp. 83--94. [Online]. Available:
  \url{https://doi.org/10.1007/978-3-031-97196-9_7}
\BIBentrySTDinterwordspacing

\bibitem{Wenzel2017_pybind11}
W.~Jakob, J.~Rhinelander, and D.~Moldovan, ``pybind11 -- seamless operability
  between c++11 and python,'' 2017, https://github.com/pybind/pybind11.

\bibitem{Kaiser2020_hpx}
H.~Kaiser, P.~Diehl, A.~Lemoine, B.~Lelbach, P.~Amini, A.~Berge,
  J.~Biddiscombe, S.~Brandt, N.~Gupta, T.~Heller, K.~Huck, Z.~Khatami,
  A.~Kheirkhahan, A.~Reverdell, S.~Shirzad, M.~Simberg, B.~Wagle, W.~Wei, and
  T.~Zhang, ``Hpx - the c++ standard library for parallelism and concurrency,''
  \emph{Journal of Open Source Software}, vol.~5, p. 2352, 09 2020.

\bibitem{Gamblin2015_spack}
T.~Gamblin, M.~LeGendre, M.~R. Collette, G.~L. Lee, A.~Moody, B.~R.
  de~Supinski, and S.~Futral, ``The spack package manager: bringing order to
  hpc software chaos,'' in \emph{SC '15: Proceedings of the International
  Conference for High Performance Computing, Networking, Storage and Analysis},
  2015, pp. 1--12.

\bibitem{Strack2026_gprat_riscv}
\BIBentryALTinterwordspacing
A.~Strack, P.~Diehl, and D.~Pfl{\"u}ger, ``{Is RISC-V Ready for Machine
  Learning? Portable Gaussian Processes Using Asynchronous Tasks},'' 2026.
  [Online]. Available: \url{https://arxiv.org/abs/2604.00736}
\BIBentrySTDinterwordspacing

\bibitem{Moellmann2026_gprat_gpu}
\BIBentryALTinterwordspacing
H.~M{\"o}llmann, D.~Pfl{\"u}ger, and A.~Strack, ``{GPU-Resident Gaussian
  Process Regression Leveraging Asynchronous Tasks with HPX},'' 2026. [Online].
  Available: \url{https://arxiv.org/abs/2602.19683}
\BIBentrySTDinterwordspacing

\bibitem{OneMath}
\BIBentryALTinterwordspacing
{UXL Foundation}. (2026) {oneAPI Math Library (oneMath)}. [Online]. Available:
  \url{https://github.com/uxlfoundation/oneMath}
\BIBentrySTDinterwordspacing

\bibitem{Strack2026_fork_join}
\BIBentryALTinterwordspacing
A.~Strack, A.~{Van Craen}, and D.~Pflüger, ``From fork-join to asynchronous
  tasks: Parallelizing tiled cholesky decomposition with openmp and hpx,''
  2026. [Online]. Available: \url{https://arxiv.org/abs/2606.11937}
\BIBentrySTDinterwordspacing

\bibitem{Abdulah2018_exageostat_starpu_cholesky}
S.~Abdulah, H.~Ltaief, Y.~Sun, M.~G. Genton, and D.~E. Keyes, ``Exageostat: A
  high performance unified software for geostatistics on manycore systems,''
  \emph{IEEE Transactions on Parallel and Distributed Systems}, vol.~29,
  no.~12, pp. 2771--2784, 2018.

\bibitem{Augonnet2011_star_pu}
C.~Augonnet, S.~Thibault, R.~Namyst, and P.-A. Wacrenier, ``{StarPU: A Unified
  Platform for Task Scheduling on Heterogeneous Multicore Architectures},''
  \emph{CCPE - Concurrency and Computation: Practice and Experience, Special
  Issue: Euro-Par 2009}, vol.~23, pp. 187--198, 02 2011.

\bibitem{Bosilca2013_parsec}
G.~Bosilca, A.~Bouteiller, A.~Danalis, M.~Faverge, T.~Herault, and J.~J.
  Dongarra, ``Parsec: Exploiting heterogeneity to enhance scalability,''
  \emph{Computing in Science \& Engineering}, vol.~15, no.~6, pp. 36--45, 2013.

\bibitem{Abdulah2022_parsec_geostat_mixed_precision}
S.~Abdulah, Q.~Cao, Y.~Pei, G.~Bosilca, J.~Dongarra, M.~G. Genton, D.~E. Keyes,
  H.~Ltaief, and Y.~Sun, ``Accelerating geostatistical modeling and prediction
  with mixed-precision computations: A high-productivity approach with
  {PaRSEC},'' \emph{IEEE Transactions on Parallel and Distributed Systems},
  vol.~33, no.~4, pp. 964--976, 2022.

\bibitem{Reguly2023_sycl_portability}
I.~Z. Reguly, ``Evaluating the performance portability of {SYCL} across {CPUs}
  and {GPUs} on bandwidth-bound applications,'' in \emph{Proceedings of the SC
  '23 Workshops of the International Conference on High Performance Computing,
  Network, Storage, and Analysis}, ser. SC-W '23.\hskip 1em plus 0.5em minus
  0.4em\relax New York, NY, USA: Association for Computing Machinery, 2023, pp.
  1038--1047.

\bibitem{Alekseenko2024_gromacs_sycl}
A.~Alekseenko, S.~P\'all, and E.~Lindahl, ``{GROMACS} on {AMD} {GPU}-based
  {HPC} platforms: Using {SYCL} for performance and portability,'' in
  \emph{Proceedings of the Cray User Group (CUG 2024)}, 2024, pp. 71--84,
  arXiv:2405.01420.

\bibitem{Breyer2022_plssvm_sycl_portability}
\BIBentryALTinterwordspacing
M.~Breyer, A.~Van~Craen, and D.~Pfl{\"u}ger, ``A comparison of sycl, opencl,
  cuda, and openmp for massively parallel support vector machine classification
  on multi-vendor hardware,'' in \emph{Proceedings of the 10th International
  Workshop on OpenCL}, ser. IWOCL '22.\hskip 1em plus 0.5em minus 0.4em\relax
  New York, NY, USA: Association for Computing Machinery, 2022. [Online].
  Available: \url{https://doi.org/10.1145/3529538.3529980}
\BIBentrySTDinterwordspacing

\bibitem{Alpay2022_adaptivecpp}
A.~Alpay, B.~Soproni, H.~W\"{u}nsche, and V.~Heuveline, ``{Exploring the
  possibility of a hipSYCL-based implementation of oneAPI},'' in
  \emph{Proceedings of the 10th International Workshop on OpenCL}, ser. IWOCL
  '22.\hskip 1em plus 0.5em minus 0.4em\relax New York, NY, USA: Association
  for Computing Machinery, 2022.

\bibitem{Godoy2023_portability_julia_kokkos}
W.~F. Godoy, P.~Valero-Lara, T.~E. Dettling, C.~Trefftz, I.~Jorquera,
  T.~Sheehy, R.~G. Miller, M.~Gonzalez-Tallada, J.~S. Vetter, and V.~Churavy,
  ``Evaluating performance and portability of high-level programming models:
  {Julia}, {Python}/{Numba}, and {Kokkos} on exascale nodes,'' in \emph{2023
  IEEE International Parallel and Distributed Processing Symposium Workshops
  (IPDPSW)}, ser. HIPS '23, 2023, pp. 373--382, arXiv:2303.06195.

\bibitem{Monil2024_matris_cholesky}
M.~A.~H. Monil, N.~R. Miniskar, P.~Valero-Lara, K.~Teranishi, and J.~S. Vetter,
  ``{MatRIS}: Addressing the challenges for portability and heterogeneity using
  tasking for matrix decomposition ({Cholesky}),'' in \emph{Asynchronous
  Many-Task Systems and Applications (WAMTA 2024)}, ser. Lecture Notes in
  Computer Science, vol. 14626.\hskip 1em plus 0.5em minus 0.4em\relax
  Springer, 2024, pp. 59--70.

\bibitem{Korakitis2022_ompss2_openacc}
O.~Korakitis, S.~G. de~Gonzalo, N.~Guidotti, J.~Barreto, J.~Monteiro, and A.~J.
  Pe{\~n}a, ``{OmpSs-2} and {OpenACC} interoperation,'' in \emph{2022 Workshop
  on Accelerator Programming Using Directives (WACCPD)}, 2022, pp. 11--21.

\bibitem{Daiss2023_hpx_kokkos_sycl}
G.~Dai\ss{}, P.~Diehl, H.~Kaiser, and D.~Pfl\"{u}~ger, ``Stellar mergers with
  hpx-kokkos and sycl: Methods of using an asynchronous many-task runtime
  system with sycl,'' in \emph{Proceedings of the 2023 International Workshop
  on OpenCL}, ser. IWOCL '23.\hskip 1em plus 0.5em minus 0.4em\relax New York,
  NY, USA: Association for Computing Machinery, 2023.

\bibitem{IntelDpcppCompiler}
\BIBentryALTinterwordspacing
{Intel Corporation}. (2025) {Intel oneAPI DPC++/C++ Compiler}. Version
  2025.1.1. [Online]. Available:
  \url{https://www.intel.com/content/www/us/en/developer/tools/oneapi/dpc-compiler.html}
\BIBentrySTDinterwordspacing

\bibitem{oneapi}
\BIBentryALTinterwordspacing
{Intel Corporation}. (2026) {oneAPI: A New Era of Heterogeneous Computing}.
  [Online]. Available:
  \url{https://www.intel.com/content/www/us/en/developer/tools/oneapi/overview.html}
\BIBentrySTDinterwordspacing

\bibitem{cublas}
{NVIDIA Corporation}, \emph{{cuBLAS} {Library} {User} {Guide}}, NVIDIA
  Corporation, 2026, [Online]. Available:
  \url{https://docs.nvidia.com/cuda/cublas/index.html}. [Accessed: Aug. 5,
  2026].

\bibitem{Pleiss2018_gpytorch_love}
\BIBentryALTinterwordspacing
G.~Pleiss, J.~R. Gardner, K.~Q. Weinberger, and A.~G. Wilson, ``Constant-time
  predictive distributions for gaussian processes,'' \emph{CoRR}, vol.
  abs/1803.06058, 2018. [Online]. Available:
  \url{http://arxiv.org/abs/1803.06058}
\BIBentrySTDinterwordspacing

\bibitem{Buttari2009_parallel_exakt_solvers}
A.~Buttari, J.~Langou, J.~Kurzak, and J.~Dongarra, ``A class of parallel tiled
  linear algebra algorithms for multicore architectures,'' \emph{Parallel
  Computing}, vol.~35, no.~1, pp. 38--53, 2009.

\bibitem{portf-stuff-optimized-for-memacc}
\BIBentryALTinterwordspacing
M.~Kowarschik and C.~Wei{\ss}, \emph{{An Overview of Cache Optimization
  Techniques and Cache-Aware Numerical Algorithms}}.\hskip 1em plus 0.5em minus
  0.4em\relax Berlin, Heidelberg: Springer Berlin Heidelberg, 2003, pp.
  213--232. [Online]. Available: \url{https://doi.org/10.1007/3-540-36574-5_10}
\BIBentrySTDinterwordspacing

\bibitem{nvidia2025_cuda}
\BIBentryALTinterwordspacing
{NVIDIA Corporation}, \emph{CUDA C++ Programming Guide}, NVIDIA, 2025, version
  12.9. [Online]. Available:
  \url{https://docs.nvidia.com/cuda/cuda-c-programming-guide/}
\BIBentrySTDinterwordspacing

\bibitem{cuda-graphs-in-cuda-10}
\BIBentryALTinterwordspacing
{Luke Durant}. (2018) {CUDA 10.0 new features \& capabilities}. [Online].
  Available: \url{https://www.nvidia.com/content/apac/gtc/ja/pdf/2018/2052.pdf}
\BIBentrySTDinterwordspacing

\bibitem{og-gprat-data-paper}
M.~Helmann, A.~Strack, and D.~Pflüger, ``{Replication Data for: GPRat:
  Gaussian Process Regression with Asynchronous Tasks},'' 2025.

\bibitem{IntelTensorFlow}
\BIBentryALTinterwordspacing
{Intel Corporation}. (2025) {Intel® Extension for TensorFlow}. Version
  2.15.0.3. [Online]. Available:
  \url{https://pypi.org/project/intel-extension-for-tensorflow/}
\BIBentrySTDinterwordspacing

\bibitem{grammarly}
{Grammarly, Inc.}, ``Grammarly,'' 2026, [Online]. Available:
  \url{https://www.grammarly.com/}. [Accessed: Aug. 5, 2026].

\bibitem{claude}
{Anthropic}, ``{Claude},'' 2026, [Online]. Available: \url{https://claude.ai}.
  [Accessed: Aug. 5, 2026].

\end{thebibliography}

\clearpage
\onecolumn
\appendices
\section{Reproducibility Information}
\label{sec:reproducibility}

A summary of the most important hardware and software properties of the test systems is provided in Tables~\ref{table:reproducibility-hardware} and \ref{table:reproducibility-software}, respectively.
The devices differ significantly in their cache layouts, memory bandwidth, and theoretical peak performance.
All specifications in Table~\ref{table:reproducibility-hardware} are taken from the respective manufacturers' public datasheets.
Whenever possible, the same software versions were chosen on all devices.
For compatibility reasons, the PyTorch and TensorFlow installations on the MI210 system have slightly lower version numbers than on the other systems.
However, GPyTorch and GPflow had the same version number on all of them.
The benchmarks were carried out using the dataset~\cite{og-gprat-data-paper} from the original GPRat publication~\cite{Helmann2026_gprat}.
The SYCL backend implementation and the benchmarking scripts used to produce all results in this work are available at \url{https://github.com/SC-SGS/GPRat}.
All measurements were obtained with GPRat release v0.4.0.

\begin{table}[H]
	\centering
	\setlength{\tabcolsep}{8pt}
	\resizebox{\textwidth}{!}{%
		\footnotesize
		\begin{tabular}{|M{2cm}|M{3.6cm}|M{3.6cm}|M{3.6cm}|M{3.6cm}|}
			\toprule
			{}  & \textbf{CPU}                                & \textbf{NVIDIA GPU}                & \textbf{AMD GPU}                   & \textbf{Intel GPU}                       \\
			\midrule
			Device    & $2\times$ AMD EPYC 9274F                    & NVIDIA A30                         & AMD Instinct MI210                 & Intel Arc B580                     \\
			\rowcolor{lightgray!50}
			Cores/CUs & $2\times$~$24$ Cores                           & 56 SMs                             & 104 CUs                            & 20 Xe Cores                        \\
			L1 cache  & \qty{64}{\kilo\byte} / core                 & \qty{192}{\kilo\byte} / SM         & \qty{16}{\kilo\byte} / CU          & \qty{256}{\kilo\byte} / Xe Core    \\
			\rowcolor{lightgray!50}
			L2 cache  & \qty{1}{\mega\byte} / core                  & \qty{24}{\mega\byte}               & \qty{8}{\mega\byte}                & \qty{18}{\mega\byte}               \\
			L3 cache  & \qty{256}{\mega\byte} / socket              & --                                 & --                                 & --                                 \\
			\rowcolor{lightgray!50}
			RAM/VRAM  & \qty{384}{\giga\byte} DDR5                  & \qty{24}{\giga\byte} HBM2e         & \qty{64}{\giga\byte} HBM2e         & \qty{12}{\giga\byte} GDDR6         \\
			Bandwidth & $2\times$~\qty{460.8}{\giga\byte\per\second} & \qty{933}{\giga\byte\per\second}   & \qty{1640}{\giga\byte\per\second}  & \qty{456}{\giga\byte\per\second}   \\
			\rowcolor{lightgray!50}
			FP64      & $2\times$~\qty{1.65}{\tera\flop\per\second} & \qty{5.161}{\tera\flop\per\second} & \qty{22.63}{\tera\flop\per\second} & \qty{1.709}{\tera\flop\per\second} \\
			\bottomrule
		\end{tabular}
	}
	\caption{Hardware specifications of the test systems.}
	\label{table:reproducibility-hardware}
	\bigskip
	\resizebox{\textwidth}{!}{%
		\begin{tabular}{|M{2cm}|M{2.6cm}|M{2.8cm}|M{2.8cm}|M{3.9cm}|M{3.5cm}|}
			\toprule
			{}   & \textbf{CPU}                                     & \textbf{NVIDIA GPU (CUDA)}                        & \textbf{NVIDIA GPU (SYCL)}                                                    & \textbf{AMD GPU}                                         & \textbf{Intel GPU}                    \\
			\midrule
			OS         & \multicolumn{4}{c|}{Ubuntu 22.04.5 LTS}          & Ubuntu 24.04.4 LTS                                                                                                                                                                                                             \\
			\rowcolor{lightgray!50}
			GPU Stack  & --                                               & CUDA 12.0.1                                       & CUDA 12.6.3                                                                   & ROCm 6.4.0                                               & Level Zero (\texttt{xe} driver) \\
			Compiler   & \texttt{gcc} 14.1.0                              & Clang 17.0.1                                      & \multicolumn{2}{c|}{Intel oneAPI DPC\texttt{++}/C++ (\texttt{icpx}) 2025.1.1} & Intel oneAPI DPC\texttt{++}/C++ (\texttt{icpx}) 2025.3.3                                   \\
			\rowcolor{lightgray!50}
			Backend    & Intel oneMKL 2024.2.1                            & cu\{BLAS, SOLVER\} 12.0.1                         & \multicolumn{3}{c|}{UXL oneMath v0.9}                                                                                                                                      \\
			Torch      & \texttt{torch==2.11.0}                           & \multicolumn{2}{c|}{\texttt{torch==2.11.0+cu126}} & \texttt{torch==2.9.1+rocm6.4}                                                 & --                                                                                         \\
			\rowcolor{lightgray!50}
			GPyTorch   & \multicolumn{4}{c|}{\texttt{gpytorch==1.13}}     & --                                                                                                                                                                                                                             \\
			TensorFlow & \multicolumn{3}{c|}{\texttt{tensorflow==2.21.0}} & \texttt{tensorflow-rocm==2.17.1}                  & --                                                                                                                                                                         \\
			\rowcolor{lightgray!50}
			GPflow     & \multicolumn{4}{c|}{\texttt{gpflow==2.9.2}}      & --                                                                                                                                                                                                                             \\
			HPX        & \multicolumn{5}{c|}{HPX 1.10.0}                                                                                                                                                                                                                                                   \\
			\rowcolor{lightgray!50}
			pybind11   & \multicolumn{5}{c|}{pybind11 2.10.3}                                                                                                                                                                                                                                              \\
			\bottomrule
		\end{tabular}%
	}
	\caption{Software specifications of the test systems.}
	\label{table:reproducibility-software}
\end{table}

\end{document}